\documentclass[aps,pra,reprint,onecolumn,notitlepage,eqsecnum,floatfix,nofootinbib,showkeys]{revtex4-2}

\usepackage{amsmath}
\usepackage{color}
\usepackage{graphicx}
\usepackage{hyperref}
\usepackage{multirow}
\usepackage{epigraph}

\DeclareMathSymbol{\mlq}{\mathord}{operators}{``}
\DeclareMathSymbol{\mrq}{\mathord}{operators}{`'}

\newcommand{\bq}{\begin{eqnarray}}
\newcommand{\eq}{\end{eqnarray}}
\newcommand{\bqn}{\begin{eqnarray*}}
\newcommand{\eqn}{\end{eqnarray*}}
\newcommand{\bqs}{\begin{subequations}}
\newcommand{\eqs}{\end{subequations}}
\newcommand{\bw}{\begin{widetext}}
\newcommand{\ew}{\end{widetext}}

\newcommand{\kk}{{\bf k}}
\newcommand{\rr}{{\bf r}}
\newcommand{\pp}{{\bf p}}

\newcommand{\DD}{{\bf D}}
\newcommand{\EE}{{\bf E}}
\newcommand{\JJ}{{\bf J}}

\def\half{\textstyle{\frac{1}{2}}}

\newcommand{\red}[1]{{#1}}

\begin{document}
\title{Sum Rules in Quantum Liquids}

\author{Riccardo Fantoni}
\email{riccardo.fantoni@scuola.istruzione.it}
\affiliation{Universit\`a di Trieste, Dipartimento di Fisica, strada
  Costiera 11, 34151 Grignano (Trieste), Italy}

\date{\today}

\begin{abstract}
We review the linear response theory in the horizontal quantum liquids framework spanning from 
Coulomb liquids to Atomic gases. \red{There are several well known references about this subject, 
like the onset of the Kramers-Kronig relations and the fluctuation-dissipation theorem. 
For the Coulomb systems we show the connection between the linear response function and the
dielectric function which settles a parallelism between statistical mechanics and 
electrostatic properties. For (very degenerate, dilute, trapped, two dimensional) Atomic (Bose) 
gases we will prove, in full generality, how the response properties of the gas depend from the 
frequency of the harmonic trap}.
\end{abstract}


\maketitle
\section{Introduction}

Linear response theory \cite{PinNoz66,MarTos84,HanMcD86} is a well known framework to describe the 
approach to thermal equilibrium in response to an external perturbation acting on a many-body 
quantum \red{or non-quantum} fluid. \red{Some recent books describing it are Refs. 
\cite{Chaikin-Lubensky,Giuliani-Vignale,Pathria-Beale}. We will define the linear response function 
in real space-time $K$ or in Fourier space-time $\chi$. We will then derive the
fluctuation-dissipation theorem relating the dissipation in the fluid, given by the imaginary
part of $\chi$, and its fluctuations, given by the van Hove dynamic response $S$. And finally
we will present the Kramers-Kronig relations as a constraint, imposed by causality, between the 
real and imaginary parts of the linear response function $\chi$.}

\red{In this manuscript we give a synthetic description of this theory and present two of her 
applications to quantum fluids: The Coulomb Liquid and an Atomic Gas. In particular, for the
Coulomb case we will introduce the dielectric function $\epsilon$ through her definition in 
terms of the linear response function $\chi$ and discuss her properties. This identification
of the liquid response with its electrostatic properties turns out to be inspiring and 
powerful. For a very degenerate, dilute, atomic Bose gas, we consider it confined to a two 
dimensional harmonic trap ``relaxing'' according to a simple, but realistic restoring force.
Using linear response theory, we then prove, in full generality, that if the harmonic 
trap has frequency $\omega_0$ the frequency of the monopole mode of the gas response 
is exactly $2\omega_0$. This has important implications when studying the properties of 
trapped atomic gases.}   

\section{Linear response theory}
\label{sec:lrt}

Let us introduce the density linear response function $K(\rr-\rr',t-t')$ for a homogeneous fluid. 
Let us indicate with $V_b$ the ``bare'' potential in vacuum.

The coupling of the fluid to the perturbing potential is described by the Hamiltonian 
\bq
H'(t)=\int d\rr \,\rho(\rr)V_b(\rr,t),
\eq
where $\rho(\rr)$ is the density operator (here we implicitly assume that the mean value of the 
density has been subtracted from $\rho(\rr)$). We will just consider the linear effect of this 
perturbation. The change in density is given by 
\bq \label{eq:deln}
\delta n(\rr,t)=\langle\rho(\rr)\rangle-\langle\rho(\rr)\rangle_0=
{\rm tr}\{[w(t)-w_0]\rho(\rr)\},
\eq
where ${\rm tr}$ denotes the trace, $w(t)=\int\psi^*(R,t)\psi(R,t)\,d^{3N}R$ is the perturbed 
density matrix whose unperturbed counterpart is $w_0=\exp(-\beta H_0)/tr\{\exp(-\beta H_0)\}$, and 
$\beta=1/k_BT$ with $k_B$ the Boltzmann constant and $T$ the absolute temperature. We are indicating 
with $\psi(R,t)$ the many-body wave function of the fluid with particles at positions
$R=(\rr_1,\rr_2,\ldots,\rr_N)$ at time $t$. This satisfies to the Schr\"odinger equation
\bq
i\hbar\frac{\partial\psi(R,t)}{\partial t}=[H_0+H'(t)]\psi(R,t),
\eq
where $H$ is the Hamiltonian of the unperturbed fluid. Then the perturbed density matrix satisfies 
to
\bq \nonumber
i\hbar\frac{\partial w(t)}{\partial t}&=&[H_0+H'(t),w(t)]\\ \label{eq:lin}
&\approx& [H_0,w(t)-w_0]+[H'(t),w_0],
\eq
where $[A,B]$ denotes the commutator $AB-BA$ and in the last step we have linearized the effect of 
the perturbation and used $[H_0,w_0]=0$. This equation is subject to the initial condition
\bq
\lim_{t\to-\infty}w(t)=w_0,
\eq
representing a state of thermal equilibrium.

The linearized equation (\ref{eq:lin}) has the following solution
\bq
w(t)-w_0=(i\hbar)^{-1}\int_{-\infty}^tdt'\,\exp\{-iH_0(t-t')/\hbar\}
[H'(t'),w_0]\exp\{iH_0(t-t')/\hbar\}.
\eq
Inserting this result into Eq. (\ref{eq:deln}) and using the cyclic invariance of the trace, 
${\rm tr}\{AB\}={\rm tr}\{BA\}$, we can write the desired result as follows
\bq \label{eq:dn}
\delta n(\rr,t)=(-i/\hbar)\int d\rr'\int_{-\infty}^t dt'\,
\langle[\rho(\rr,t),\rho(\rr',t')]\rangle_0 V_b(\rr',t).
\eq
Again the angle parenthesis $\langle A\rangle_0={\rm tr}\{w_0A\}$ denotes the mean value on the 
equilibrium state and $\rho(\rr,t)$ is the Heisenberg operator
\bq \label{eq:Heis}
\rho(\rr,t)=\exp(iH_0t/\hbar)\rho(\rr)\exp(-iH_0t/\hbar).
\eq
So
\bq \label{eq:K}
K(\rr-\rr',t-t')=(-i/\hbar)\theta(t-t')\langle[\rho(\rr,t),\rho(\rr',t')]
\rangle_0.
\eq
This result clearly embodies the causality property through the Heaviside step function $\theta$.

Introducing the notation
\bq
\chi''(k,t-t')=(1/2\hbar)\int d(\rr-\rr')\,\exp[-i\kk\cdot(\rr-\rr')]
\langle[\rho(\rr,t),\rho(\rr',t')]\rangle_0,
\eq
we see, from Eq. (\ref{eq:K}) that the Fourier transform of $K$ is the convolution integral of the 
Fourier transform of $\chi''(k,t)$, that we will indicate with $\chi''(k,\omega)$, and of the 
Heaviside step function, that is equal to $i/(\omega+i\eta)$ with $\eta$ a small positive quantity.
We can then write the space-time Fourier transform of $K$ like so
\bq \label{eq:chi}
\chi(k,\omega)=-\int_{-\infty}^\infty \frac{d\omega'}{\pi}\,
\chi''(k,\omega')/(\omega-\omega'+i\eta).
\eq
Using the rule $(\omega+i\eta)^{-1}=P(1/\omega)-i\pi\delta(\omega)$, where $P$ denotes the Cauchy 
principal part, this can be written like so
\bq
\chi(k,\omega)=-P\int_{-\infty}^\infty \frac{d\omega'}{\pi}\,
\chi''(k,\omega')/(\omega-\omega')+i\chi''(k,\omega).
\eq
Since $\chi''(k,t)$ is written in terms of the commutator of Hermitian operators it can be readily 
shown that $\chi''(k,\omega)$ must be real. So we can write 
\bq \label{eq:cpp}
{\rm Im}\chi(k,\omega)=\chi''(k,\omega).
\eq

\section{Fluctuation-dissipation theorem}

We now worry about the relationship between the density response function and the van Hove dynamic 
response $S(\kk,\omega)$. Let us define the autocorrelation density function as
\bq
G(\rr-\rr',t-t')=\frac{1}{n}\langle\rho(\rr,t)\rho(\rr',t')\rangle_0,
\eq
whose space-time Fourier transform is $S(\kk,\omega)$. The connection between $G$ e $K$ that gush 
from Eq. (\ref{eq:K}) can be rewritten in Fourier transform like so
\bq
\chi(\kk,\omega)=(n/\hbar)\int_{-\infty}^\infty\frac{d\omega'}{2\pi}
[S(\kk,\omega)-S(-\kk,-\omega)]/(\omega-\omega'+i\eta).
\eq
This has the same form of Eq. (\ref{eq:chi}) so that
\bq \label{eq:imc}
{\rm Im}\chi(\kk,\omega)=(-n/2\hbar)[S(\kk,\omega)-S(-\kk,-\omega)].
\eq

For a fluid in thermodynamic equilibrium we must have
\bq
S(-\kk,-\omega)=\exp(-\hbar\beta\omega)S(\kk,\omega).
\eq
In order to prove this property we observe that its inverse space-time Fourier transform reads 
\bq \label{eq:G}
G(-\rr,-t)=\exp\left(-i\hbar\beta\frac{\partial}{\partial t}\right)G(\rr,t),
\eq
since under time Fourier transform $\partial/\partial t\to -i\omega$. But Eq. (\ref{eq:G}) can 
readily be proven through the following steps (where, once again we use the cyclic invariance of the 
trace and the definition of the Heisenberg operator, Eq. (\ref{eq:Heis}))
\bq \nonumber
{\rm tr}\{\exp(-\beta H_0)\rho({\bf 0},0)\rho(\rr,t)\}&=&
{\rm tr}\{\rho(\rr,t)\exp(-\beta H_0)\rho({\bf 0},0)\}\\ \nonumber
&=&{\rm tr}\{\exp(-\beta H_0)\rho(\rr,t-i\hbar\beta)\rho({\bf 0},0)\}\\
&=&\exp(-i\hbar\beta\partial/\partial t)
{\rm tr}\{\exp(-\beta H_0)\rho(\rr,t)\rho({\bf 0},0)\}.
\eq

In the classical limit, for $\beta$ small, Eq. (\ref{eq:imc}) becomes
\bq
{\rm Im}\chi(\kk,\omega)=(-n\beta\omega/2)S(\kk,\omega).
\eq

\section{Kramers-Kronig relations}

Causality imposes that the response function $K(\rr,t)$ vanish for $t<0$. In other words the fluid 
is influenced only by the action of the external perturbation in the past. Introducing the 
``intermediate'' response function $\chi(\kk,t)$ as the space Fourier transform of $K(\rr,t)$, we 
have
\bq \label{eq:causality}
\chi(\kk,t)=0~~~{\rm for}~~~t<0.
\eq
On the other hand
\bq
\chi(\kk,t)=\int_{-\infty}^\infty\frac{d\omega}{2\pi}\exp(-i\omega t)
\chi(\kk,\omega).
\eq
Extending the definition of $\chi(\kk,\omega)$ from real to complex frequencies, we can calculate 
this integral through contour methods and for $t<0$ we can close the contour with the semicircle at 
infinity above the real axis. The contribution from the integration on the semicircle vanishes since 
$\chi(\kk,\omega)\propto\omega^{-2}$ at high frequency. So the causality property 
(\ref{eq:causality}) is guaranteed if $\chi(\kk,\omega)$ is analytic in the upper part of the 
complex frequency plane.

\begin{figure}[hbt]
\begin{center}
\includegraphics[width=9cm]{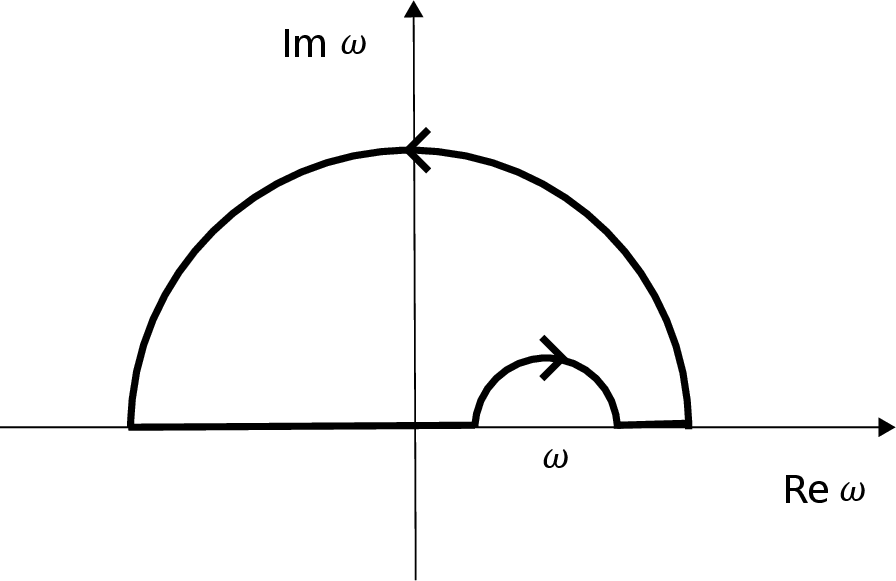}
\end{center}
\caption[]{Integration contour on the complex $\omega$ plane.}
\label{fig:contorno}
\end{figure}

Let us now consider the integral
\bq
\oint\frac{\chi(\kk,\omega')}{\omega-\omega'}\,d\omega'=0,
\eq
on the contour $\Gamma$ shown in Fig. \ref{fig:contorno}. This contour integral vanishes due to the 
analiticity of $\chi(\kk,\omega)$. The contribution from the semicircle at infinity is again zero, 
so that
\bq
P\int_{-\infty}^\infty d\omega'\,\frac{\chi(\kk,\omega')}{\omega'-\omega}
-i\pi\chi(\kk,\omega)=0,
\eq
where again $P$ denotes the Cauchy principal part of the integral on the real frequency axis and the 
second term comes from the integration over the small semicircle around the point $\omega$. If we 
now separate $\chi(\kk,\omega)$ into its real and imaginary parts we find
\bq
P\int_{-\infty}^\infty d\omega'\,\frac{{\rm Re}\chi(\kk,\omega')}{\omega'-\omega}
+\pi{\rm Im}\chi(\kk,\omega)=0,
\eq
and
\bq
P\int_{-\infty}^\infty d\omega'\,\frac{{\rm Im}\chi(\kk,\omega')}{\omega'-\omega}
-\pi{\rm Re}\chi(\kk,\omega)=0.
\eq
These are the Kramers-Kronig relations.
%

\section{The dielectric function}

In a Coulomb liquid \red{(either in its quantum version, the {\sl Jellium}
\cite{Fantoni95b,Fantoni13g,Fantoni18c,Fantoni21b,Fantoni21d,Fantoni21i} 
or in its non-quantum version, the {\sl one-component-plasma},
\cite{Fantoni95a,Fantoni03a,Fantoni08c,Fantoni12b,Fantoni13e,Fantoni13f,Fantoni16b,Fantoni19a,Fantoni23a})}, 
the connection with the longitudinal dielectric function $\epsilon(k,\omega)$, 
becomes apparent from the Poisson equations
\bq
\nabla\cdot\DD(\rr,t)&=&-4\pi e n_e(\rr,t),\\
\nabla\cdot\EE(\rr,t)&=&-4\pi e [n_e(\rr,t)+\delta n(\rr,t)], 
\eq
which yield
\bq
\frac{1}{\epsilon(k,\omega)}=\frac{\kk\cdot\EE(\kk,\omega)}{\kk\cdot\DD(\kk,\omega)}
=1+\frac{\delta n(\kk,\omega)}{n_e(\kk,\omega)}=1+\frac{4\pi e^2}{k^2}\chi(k,\omega),
\eq
since from Eqs. (\ref{eq:dn}) and (\ref{eq:K}) follows 
$\delta n(\kk,\omega)=\chi(k,\omega)V_b(\kk,\omega)$ where $\chi(k,\omega)$ is the Fourier transform 
of $K(|\rr-\rr'|,t-t')$ and
\bq
V_b(\kk,\omega)=\frac{4\pi e^2}{k^2}n_e(\kk,\omega).
\eq
Of course the field $\EE$ and the associated screened or ``Hartree'' potential 
$V_H(\kk,\omega)=V_b(\kk,\omega)/\epsilon(k,\omega)$ would be experienced by a second test charge 
introduced into the plasma, rather than by the particles of the plasma. The latter also experience 
effects which involve the precise ``hole'' a particle of the plasma digs around itself. This latter 
effect brings about the so called local field corrections.

In addition to $\chi(\kk,\omega)$ which relates the displaced charge density to the potential {\sl 
in vacuo}, it is useful to introduce yet another longitudinal response function, 
$\tilde{\chi}(k,\omega)$ say, by exploiting further the analogy with elementary electrostatics. This 
relates $n(\kk,\omega)$ directly to the Hartree potential through
\bq
n(\kk,\omega)=\tilde{\chi}(k,\omega)V_H(\kk,\omega).
\eq 
We then have
\bq
\epsilon(k,\omega)=1-\frac{4\pi e^2}{k^2}\tilde{\chi}(k,\omega).
\eq
The expression $\chi(k,\omega)=\tilde{\chi}(k,\omega)/\epsilon(k,\omega)$ accounts at one stroke for the 
{\sl long range} effects of the Coulomb interactions (the resonance at the plasma frequency, determined 
by $\epsilon(k,\omega)=0$, is brought about explicitly in the denominator.

The simplest useful approximation to the dielectric function of the plasma is obtained by approximating 
$\tilde{\chi}$ by the density response function of an ideal gas. This corresponds to the Vlasov theory 
for the classical plasma and to the Lindhard theory  for the degenerate electron fluid. Refinements of 
these theories aims at incorporating the effects of ``exchange and correlation'' in $\tilde{\chi}$. This 
expression being an abbreviation for the short range effects arising from the statistics (``exchange'') 
and long range effect arising from the Coulomb interaction (``correlation''). Of course the exchange 
effects are absent in the classical limit. 

\section{Very degenerate atomic gases}

\noindent\fbox{%
    \parbox{\textwidth}{%
{\sl 
Consider a dilute (Bose) gas with interaction energy per unit area equal to $U_0$, confined to a 2D 
isotropic harmonic well with harmonic frequency $\omega_0$. We will show that the kinetic and 
potential energy separately can be written in the general form 
$(T,V)=\hbar\omega_0 f_{(T,V)}(\alpha)$ where $\alpha=NU_0m/\hbar^2$ is independent of $\omega_0$. 
We will also show that the mean square radius of the cloud can be written in the form 
$d^2Ng(\alpha)$.
}
}}\vspace{.7cm}
\mbox{}
In the Hartree approximation, the ground state energy of the system is given by a 
Ginzburg-Pitaevskii-Gross energy functional
\bq
E_0[\psi(\rr)]=\int d^2\rr\left[\frac{\hbar^2}{2m}|\nabla\psi|^2+\frac{m\omega_0^2}{2}r^2|\psi|^2+\frac{U_0}{2}|\psi|^4\right],
\eq
where $\rr=(x,y)$ and $\psi(\rr)$ is the order parameter.

The total ground state energy $E_0$ and wavefunction are determined minimizing $E_0[\psi]$ with the 
constraint
\bq
\int |\psi(\rr)|\,d^2\rr=N,
\eq
where $N$ is the number of particles in the gas.

Let's introduce the root mean square zero point displacement $d$
\bq
d=\sqrt{\frac{\hbar}{m\omega_0}},
\eq
and let's make the following rescaling
\bq
\rr_1=\rr/d,\\
\psi=f\frac{\sqrt{N}}{d}.
\eq
The energy functional becomes
\bq
E_0&=&d^2\int d^2\rr_1\left[\frac{\hbar^2}{2m}\left|\frac{\nabla_1f}{d}\frac{\sqrt{N}}{d}\right|^2+\frac{m\omega_0^2d^2}{2}r_1^2\left|\frac{\sqrt{N}}{d}f\right|^2+\frac{U_0}{2}\left|\frac{\sqrt{N}}{d}f\right|^4\right]\\
&=&N\frac{\hbar\omega_0}{2}\int d^2\rr_1[|\nabla_1 f|^2+r_1^2|f|^2+\alpha |f|^4],
\eq
where $\alpha=NU_0m/\hbar^2$.

The order parameter normalization becomes,
\bq
N=\int d^2\rr |\psi|^2=d^2\int d^2\rr_1\left|f\frac{\sqrt{N}}{d}\right|^2=N\int d^2\rr_1|f|^2,
\eq
or $\int d^2\rr_1 |f|^2=1$.

Now to find the total ground state energy we need to minimize
\bq
\delta\left\{\int d^2\rr_1[|\nabla_1 f|^2+r_1^2|f|^2+\alpha |f|^4]\right\}=0,
\eq
with the constraint $\int d^2\rr_1 |f|^2=1$.

The function $f_0$ that solves the problem can only be a function of $\rr_1$ and $\alpha$, i.e. 
$f_0=f_0(\rr_1,\alpha)$.

So we can say
\bq \label{eq:opE}
\frac{E_0}{N}&=&\hbar\omega_0\frac{1}{2}\int d^2\rr_1[|\nabla_1 f_0|^2+r_1^2|f_0|^2+\alpha |f_0|^4]
=\hbar\omega_0 f_E(\alpha),\\
\frac{V}{N}&=&\hbar\omega_0\frac{1}{2}\int d^2\rr_1r_1^2|f_0|^2
=\hbar\omega_0 f_V(\alpha),\\
\frac{U}{N}&=&\hbar\omega_0\frac{1}{2}\int d^2\rr_1\alpha |f_0|^4
=\hbar\omega_0 f_U(\alpha),\\ \label{eq:opT}
\frac{T}{N}&=&\hbar\omega_0\frac{1}{2}\int d^2\rr_1|\nabla_1 f_0|^2
=\hbar\omega_0 f_T(\alpha).
\eq
The mean square radius of the cloud is
\bq
\langle r^2\rangle=\int d^2\rr\,r^2|\psi_0|^2=d^2N\int d^2\rr_1\,r_1^2|f_0|^2=d^2Ng(\alpha).
\eq

\noindent\fbox{%
    \parbox{\textwidth}{%
{\sl 
We consider a ``relaxing'' simple harmonic oscillator, i.e. one in which the restoring force 
$F(t)$ is related to the displacement $x(t')$ for $t'<t$ by the formula
\bq
F(t)=\int K(t-t')x(t')\,dt',
\eq
with $K(t-t')$ given by
\bq
K(t-t')=\left\{\begin{array}{ll}
k_s\delta(t-t')-(k_s-k_t)\tau^{-1}\exp[-(t-t')/\tau] & t>t'\\
0 & t<t'
\end{array}\right..
\eq
We find next the linear response function of the oscillator as a function of $\omega_t^2=k_t/m$, 
$\omega_s^2=k_s/m$, and $\tau$, and find the damping of the oscillations in the limit 
$\omega_s\gtrsim\omega_t\gg\tau^{-1}$ and $\omega_s\gtrsim\omega_t\ll\tau^{-1}$.
}
}}\vspace{.7cm}
\mbox{}
Imagine to perturb the system with a force $G(t)$, the equation of motion of the harmonic oscillator 
will look like
\bq
m\ddot{x}(t)+F(t)=G(t).
\eq
The linear response function $\chi(\omega)$ is defined as
\bq
\chi(\omega)=\frac{\tilde{x}(\omega)}{\tilde{G}(\omega)},
\eq
where we indicate with a tilde the time Fourier transform of the corresponding function. Taking the 
time Fourier transform of the equation of motion we get
\bq
\tilde{G}=-\omega^2\tilde{x}+\omega_s^2\tilde{x}-\frac{\omega_s^2-\omega_t^2}{\tau}\tilde{x}\int_{-\infty}^\infty\theta(t)e^{-t/\tau}e^{i\omega t}\,dt,
\eq
where $\omega_s^2=k_s/m$, $\omega_t^2=k_t/m$, and we used the property of the Fourier transform to 
transform a convolution into a product. 

Now we have
\bq
\int_{-\infty}^\infty \theta(t)e^{-t/\tau}e^{i\omega t}\,dt=\int_0^\infty e^{-t\left(\frac{1}{\tau}-i\omega\right)}\,dt=\frac{1}{\frac{1}{\tau}-i\omega},
\eq
since $\tau>0$.

So our equation of motion becomes
\bq
\tilde{G}=-\omega^2\tilde{x}+\omega_s^2\tilde{x}-\frac{\omega_s^2-\omega_t^2}{1-i\omega\tau}\tilde{x},
\eq
and the linear response function looks like
\bq
\chi(\omega)=\frac{\tilde{x}(\omega)}{\tilde{G}(\omega)}=\frac{1}{-\omega^2+\omega_s^2-\frac{\omega_s^2-\omega_t^2}{1-i\omega\tau}}
\eq
Introducing adimensional frequencies $\bar{\omega}=\omega\tau$, 
$\bar{\omega}_{s,t}=\omega_{s,t}\tau$ we get
\bq
\chi(\bar{\omega})=-\tau^2\left(\frac{\bar{\omega}+i}{\bar{\omega}^3+i\bar{\omega}^2-\bar{\omega}\bar{\omega}_s^2-i\bar{\omega}^2_t}\right).
\eq
To find the damping of the oscillations we need to find the poles of $\chi(\bar{\omega})$.

Note that 
\bq
x(t)=\int_{-\infty}^\infty\chi(t-t')G(t')\,dt'
\eq
this gives us two informations about the poles of $\chi(\omega)$ extended from real to complex 
frequencies:
\begin{itemize}
\item[i.] Causality implies that $\chi(t)=0$ for $t<0$, i.e.
\bq
\int_{-\infty}^\infty\frac{d\omega}{2\pi}\,e^{-i\omega t}\chi(\omega)=0~~~\mbox{for $t<0$},
\eq
which means that $\chi(\omega)$ cannot have poles in the upper half of the complex frequency plane (has to be analytic there).
\item[ii.] Since $\chi(t)$ and $G(t)$ are physical quantities, hence real, 
$\chi(-\omega^*)=\chi^*(\omega)$.
\end{itemize}
Now, since the denominator of $\chi(\omega)$ is a cubic, it has to have 3 roots. Because of i. and ii. these 3 roots have to be of the following kind
\bq \nonumber
\bar{\omega}_1&=&-ia,\\ \nonumber
\bar{\omega}_2&=&-\bar{\omega}_3^*=-ic+b,
\eq
with $a,b,c$ three real positive numbers. We want to find the damping of the oscillations produced 
by the term $b$ in $\chi(t)$. That is, we want to find $c$ (the $\omega=\bar{\omega}_1/\tau$ pole 
will give rise to a dumped non oscillating factor instead).

So we want
\bq \nonumber
&&\bar{\omega}^3+i\bar{\omega}^2-\bar{\omega}\bar{\omega}_s^2-i\bar{\omega}^2_t\\ \nonumber
&=&(\bar{\omega}+ia)(\bar{\omega}+ic+b)(\bar{\omega}+ic-b)\\ \nonumber
&=&(\bar{\omega}+ia)(\bar{\omega}^2-c^2+2ic\bar{\omega}-b^2)\\ \nonumber
&=&\bar{\omega}^3+i\bar{\omega}^2(2c+a)-\bar{\omega}(b^2+c^2+2ca)-i(b^2+c^2)a,
\eq
which gives
\bq \label{eq:ls}
\left\{\begin{array}{l}
2c+a=1\\
(b^2+c^2)+2ca=\bar{\omega}_s^2\\
(b^2+c^2)a=\bar{\omega}_t^2
\end{array}\right.
\eq
This linear system reduces to
\bq \label{eq:c}
c^3-c^2+c\left(\frac{1+\bar{\omega}_s^2}{4}\right)-\left(\frac{\bar{\omega}_s^2-\bar{\omega}_t^2}{8}\right)=0
\eq
Now for small $\tau$ Eq. (\ref{eq:c}) reduces to 
\bq
c^3-c^2+c/4\approx 0,
\eq
which has solutions $c=0,a=1,b=\bar{\omega}_t$ and $c=1/2,a=0,b=\sqrt{\bar{\omega}_s^2-1/4}$. Both these solutions are uninteresting since the first one gives
\bq
\chi(\bar{\omega})\approx -\tau^2\frac{1}{(\bar{\omega}+\bar{\omega}_t)(\bar{\omega}-\bar{\omega}_t)},
\eq
which corresponds to an undamped 
\bq
\chi(t)\approx(\tau/2\omega_t)\sin(\omega_t t).
\eq
While the second one gives a complex $b$, since $\bar{\omega}_s^2\ll 1/4$, which cannot be.\\
We then learn that for the case $\bar{\omega}_s^2\gtrsim\bar{\omega}_t^2\ll 1$ we need to expand 
around $c=0$ to obtain
\bq
c\left(\frac{1}{4}\right)-\left(\frac{\bar{\omega}_s^2-\bar{\omega}_t^2}{8}\right)\approx 0,
\eq
with solution
\bq
c\approx\frac{\bar{\omega}_s^2-\bar{\omega}_t^2}{2},
\eq
and $a\approx 1$ and $b\approx\bar{\omega}_s$. So we have
\bq
\chi(\bar{\omega})\approx -\tau^2\frac{1}{\left[\bar{\omega}+i\left(\frac{\bar{\omega}_s^2-\bar{\omega}_t^2}{2}\right)+\bar{\omega}_s\right]\left[\bar{\omega}+i\left(\frac{\bar{\omega}_s^2-\bar{\omega}_t^2}{2}\right)-\bar{\omega}_s\right]},
\eq
which corresponds to
\bq
\chi(t)\approx\tau e^{-\left(\frac{\omega_s^2-\omega_t^2}{2}\right)\tau t}\frac{\sin(\omega_s t)}{\omega_s}.
\eq
 
For large $\tau$ Eq. (\ref{eq:c}) reduces to
\bq
c\left(\frac{\bar{\omega}_s^2}{4}\right)-\left(\frac{\bar{\omega}_s^2-\bar{\omega}_t^2}{8}\right)\approx 0,
\eq
which gives
\bq
c\approx\frac{1}{2}\left(\frac{\bar{\omega}_s^2-\bar{\omega}_t^2}{\bar{\omega}_s^2}\right).
\eq
Now for $\bar{\omega}_s^2\gtrsim\bar{\omega}_t^2\gg 1$ $c$ is small and from Eq. (\ref{eq:ls})
we find $a\approx 1$ and $b\approx\bar{\omega}_s$. So we have
\bq
\chi(\bar{\omega})\approx -\tau^2\frac{1}{\left[\bar{\omega}+i\left(\frac{\bar{\omega}_s^2-\bar{\omega}_t^2}{2\bar{\omega}_s^2}\right)+\bar{\omega}_s\right]\left[\bar{\omega}+i\left(\frac{\bar{\omega}_s^2-\bar{\omega}_t^2}{2\bar{\omega}_s^2}\right)-\bar{\omega}_s\right]},
\eq
which corresponds to
\bq
\chi(t)\approx\tau e^{-\left(\frac{\omega_s^2-\omega_t^2}{2\omega_s^2}\right)t/\tau}\frac{\sin(\omega_s t)}{\omega_s}.
\eq

\noindent\fbox{%
    \parbox{\textwidth}{%
{\sl 
Consider a (boson) system living in a plane with Hamiltonian
\bq
\hat{H}_0&=&\sum_{i=1}^N\left[\frac{\hat{\pp}_i^2}{2m}+\hat{V}_{ext}(\hat{\rr_i})\right]+
\frac{U_0}{2}\sum_{i,j=1}^N\delta^2(\hat{\rr}_i-\hat{\rr}_j),\\
\hat{V}_{ext}&=&\frac{1}{2}m\omega_0^2\sum_{i=1}^N\hat{\rr}_i^2.
\eq
We can show that for this system the frequency of the monopole mode is exactly $2\omega_0$ 
irrespective of the value of $\alpha=NU_0m/\hbar^2$.
}
}}\vspace{.7cm}
\mbox{}
Since we are in a plane $\hat{\rr}_i=(\hat{r}_{i_1},\hat{r}_{i_2})$ and 
$\hat{\pp}_i=(\hat{p}_{i_1},\hat{p}_{i_2})$. It is useful to introduce the following two operators
\bq \label{eq:Q}
\hat{Q}_\alpha&=&\sum_{i=1}^N\hat{r}_{i_\alpha}^2,\\ \label{eq:Lam}
\hat{\Lambda}_\alpha&=&\frac{1}{2m}\sum_{i=1}^N[\hat{p}_{i_\alpha}\hat{r}_{i_\alpha}+\hat{r}_{i_\alpha}\hat{p}_{i_\alpha}],
\eq
for each dimension $\alpha=1,2$ and the usual density and current operators
\bq \label{eq:rho}
\hat{\rho}(\rr)&=&\sum_{i=1}^N\delta(\rr-\hat{\rr}_i),\\ \label{eq:J}
\hat{\JJ}(\rr)&=&\frac{1}{2m}\sum_{i=1}^N[\hat{\pp}_i\delta(\rr-\hat{\rr}_i)+\delta(\rr-\hat{\rr}_i)\hat{\pp}_i],
\eq
So that for example
\bq
\hat{V}_{ext}=\frac{1}{2}m\omega_0^2\sum_{\alpha=1}^2\hat{Q}_\alpha.
\eq
This {\sl harmonic trap} may be perturbed by a harmonic perturbation
$\hat{H}'=\lambda\sum_{\alpha=1}^2\hat{Q}_\alpha$.
This we will do next.

\subsection{Moments Sum Rules}

Imagine we apply to the system $\hat{H}_0$ a perturbation
\bq \label{eq:Hp}
\hat{H}'(t)=\sum_{\alpha=1}^2\int d^2\rr\,\hat{A}_\alpha(\rr)\lambda_\alpha(\rr,t),
\eq
where $\lambda_\alpha(\rr,t)$ are external fields and $\hat{A}_\alpha(\rr)$ are observables of the system coupled to the fields. Our system now is $\hat{H}=\hat{H}_0+\hat{H}'$. We can then apply the formalism developed in Sec. \ref{sec:lrt} for the linear response theory.

Let us consider the moments of the dissipation spectrum, defined as 
\bq \label{eq:Mnd}
M_{\alpha\beta}^n(k)=-\int_{-\infty}^\infty\frac{d\omega}{2\pi}\,\omega^n\,{\rm Im}\chi_{\alpha\beta}(k,\omega).
\eq
Since ${\rm Im}\chi_{\alpha\beta}(k,\omega)$ is an odd function of $\omega$ due to Eq. 
(\ref{eq:imc}), the even moments vanish. We will just be interested in the first two non-zero 
moments $n=1,3$. The moments are related to the fluctuations of the observables 
$\hat{A}_\alpha(\rr)$ by the fluctuation-dissipation theorem in the following way
\bq
M_{\alpha\beta}^n(k)&=&-\frac{2}{(-i)^n}\left.\frac{\partial^n\chi_{\alpha\beta}''(k,t)}{\partial t^n}\right|_{t=0},\\
\chi_{\alpha\beta}''(k,t)&=&-\frac{1}{2\hbar}\langle\left[\hat{A}_\alpha(\kk,t),\hat{A}_\beta(-\kk,0)\right]\rangle_0,
\eq
where we used Eq. (\ref{eq:cpp}), $\langle\ldots\rangle_0$ denotes the expectation value over the 
ground state of the unperturbed system $\hat{H}_0$, $[\ldots]$ is the commutator, 
$\hat{A}_\alpha(\kk,t)$ is the Heisenberg representation of the spatial Fourier transform of 
$\hat{A}_\alpha(\rr)$ such that 
\bq
\frac{\partial\hat{A}(t)}{\partial t}=[\hat{A}(t),\hat{H}_0]/i\hbar.
\eq

Let us write down the first two non-zero ones
\bq \label{eq:M1d}
M_{\alpha\beta}^1(k)&=&\frac{1}{\hbar^2}\langle\left[\left[\hat{A}_\alpha,\hat{H}_0\right],\hat{A}_\beta\right]\rangle_0,\\ \label{eq:M3d}
M_{\alpha\beta}^3(k)&=&\frac{1}{\hbar^4}\langle\left[\left[\left[\left[\hat{A}_\alpha,\hat{H}_0\right],\hat{H}_0\right],\hat{H}_0\right],\hat{A}_\beta\right]\rangle_0,
\eq
where all the $\hat{A}$ observables are taken at $(k,0)$.

Let us now choose, in Eq. (\ref{eq:Hp}),the external field independent of time, space, and index 
$\alpha$, and
\bq
\lambda_\alpha(\rr,t)&=&\lambda=\mbox{constant},\\
\hat{A}_\alpha(\rr)&=&r_\alpha^2\hat{\rho}(\rr),
\eq
with $\hat{\rho}$ defined in Eq. (\ref{eq:rho}). We can then evaluate the first and third moments at 
$k=0$. Let us start with $M^1_{\alpha\beta}(k=0)$. Proceeding step by step, we find
\bq \label{eq:QH}
\left[\hat{Q}_\alpha,\hat{H}_0\right]&=&2i\hbar\hat{\Lambda}_\alpha,\\ \label{eq:QLam}
\left[\hat{Q}_\beta,\left[\hat{Q}_\alpha,\hat{H}_0\right]\right]&=&
\left[\hat{Q}_\beta,2i\hbar\hat{\Lambda}_\alpha\right]=-\frac{4\hbar^2}{m}\delta_{\alpha\beta}\hat{Q}_\alpha,
\eq
where $\hat{Q}_\alpha$ and $\hat{\Lambda}_\alpha$ are defined in Eqs. (\ref{eq:Q}) and 
(\ref{eq:Lam}) and we used the usual commutation relations 
$[\hat{r}_{i_\alpha},\hat{r}_{j_\beta}]=[\hat{p}_{i_\alpha},\hat{p}_{j_\beta}]=0, [\hat{r}_{i_\alpha},\hat{p}_{j_\beta}]=i\hbar\delta_{ij}\delta_{\alpha\beta}$, so that only the kinetic
energy term in $\hat{H}_0$ contributes to the first moment.

So using Eq. (\ref{eq:QLam}) into Eq. (\ref{eq:M1d}) we find at $k=0$
\bq \label{eq:M1}
M^1_{\alpha\beta}(0)=\frac{4}{m}\delta_{\alpha\beta}\langle\hat{Q}_\alpha\rangle_0.
\eq

In order to find the third moment is convenient to rearrange the commutators in Eq. (\ref{eq:M3d}) 
as follows
\bq 
M_{\alpha\beta}^3(k)&=&\frac{1}{\hbar^4}\langle\left[\left[\hat{A}_\beta,\hat{H}_0\right],\left[\left[\hat{A}_\alpha,\hat{H}_0\right],\hat{H}_0\right]\right]\rangle_0,
\eq
so that at $k=0$ one finds from Eq. (\ref{eq:QH})
\bq \label{eq:M3p}
M_{\alpha\beta}^3(0)&=&-\frac{4}{\hbar^2}\langle\left[\hat{\Lambda}_\beta,\left[\hat{\Lambda}_\alpha,\hat{H}_0\right]\right]\rangle_0.
\eq
Let us start by calculating the commutator $\left[\hat{\Lambda}_\alpha,\hat{U}\right]$ where
\bq \label{eq:U}
\hat{U}&=&\frac{U_0}{2}\int d^2\rr\,\hat{\rho}^2(\rr)=\frac{U_0}{2}\sum_{i,j=1}^N\delta^2(\hat{\rr}_i-\hat{\rr}_j),\\
\hat{\Lambda}_\alpha &=&\int d^2\rr\,r_\alpha\hat{J}_\alpha(\rr).
\eq
From Eq. (\ref{eq:app}) in Appendix we get
\bq
\left[\hat{\Lambda}_\alpha,\hat{U}\right]=\frac{i\hbar}{m}U_0\int d^2\rr\,\hat{\rho}(\rr)\nabla_\alpha\left[r_\alpha\hat{\rho}(\rr)\right].
\eq
Then, from the identity
\bq
\hat{\rho}(\rr)\nabla_\alpha\left[r_\alpha\hat{\rho}(\rr)\right]=\frac{1}{2}
\{\hat{\rho}^2(\rr)+\nabla_\alpha\left[r_\alpha\hat{\rho}^2(\rr)\right]\},
\eq
and from the boundedness of the system follows
\bq
\left[\hat{\Lambda}_\alpha,\hat{U}\right]=\frac{i\hbar}{m}\frac{U_0}{2}\int d^2\rr\,\hat{\rho}^2(\rr)=\frac{i\hbar}{m}\hat{U},
\eq 
where in the last equality we used Eq. (\ref{eq:U}).

We have already calculated the commutator with the external potential
\bq \nonumber
\left[\hat{\Lambda}_\alpha,\hat{V}_{ext}\right]&=&
\left[\hat{\Lambda}_\alpha,\half m\omega^2_0\sum_\beta\hat{Q}_\beta\right]\\ \nonumber
&=&-\frac{2i\hbar}{m}\left(\half m\omega_0^2\sum_\beta\hat{Q}_\beta\delta_{\alpha\beta}\right)\\
&=&-\frac{2i\hbar}{m}\hat{V}_\alpha
\eq
where in the second equality we used Eq. (\ref{eq:QLam}) and the last equality defines 
$\hat{V}_\alpha=\frac{1}{2}m\omega_0^2\sum_i\hat{r}^2_{i_\alpha}$.

All that is left is to calculate the commutator $\left[\hat{\Lambda}_\alpha,\hat{K}\right]$
where $\hat{K}=\sum_i\hat{\pp}_i^2/2m$ is the kinetic energy operator. So now we get
\bq
\left[\hat{\Lambda}_\alpha,\hat{K}\right]=\frac{i\hbar}{m^2}\sum_i\hat{p}^2_{i_\alpha}=
\frac{2i\hbar}{m}\sum_i\hat{K}_\alpha,
\eq
where in the last equality we have defined $\hat{K}_\alpha=\sum_i\hat{p}_{i_\alpha}^2/2m$.

So, collecting all the pieces
\bq \label{eq:Lam-H}
\left[\hat{\Lambda}_\alpha,\hat{H}_0\right]=\frac{i\hbar}{m}
\left\{2\hat{K}_\alpha-2\hat{V}_\alpha+\hat{U}\right\}.
\eq
Now taking the ground state expectation value of this commutator one gets the virial theorem since 
$\langle\left[\hat{\Lambda}_\alpha,\hat{H}_0\right]\rangle_0=
\langle\hat{\Lambda}_\alpha\rangle_0E-E\langle\hat{\Lambda}_\alpha\rangle_0=0$, namely
\bq \label{eq:vt}
2\langle\hat{K}_\alpha\rangle_0-2\langle\hat{V}_\alpha\rangle_0+\langle\hat{U}\rangle_0=0.
\eq

We can now determine the last necessary commutator (using Eq. (\ref{eq:Lam-H}))
\bq
\left[\hat{\Lambda}_\beta,\left[\hat{\Lambda}_\alpha,\hat{H}_0\right]\right]&=&
\frac{i\hbar}{m}\left[\hat{\Lambda}_\beta,
2\hat{K}_\alpha-2\hat{V}_\alpha+\hat{U}\right]\\
&=&\left(\frac{i\hbar}{m}\right)^2\left\{4\hat{K}_\alpha\delta_{\alpha\beta}+4\hat{V}_\alpha\delta_{\alpha\beta}+\hat{U}\right\}.
\eq
So that in the end, using this result into Eq. (\ref{eq:M3p}), we find for the third moment
\bq \label{eq:M3}
M^3_{\alpha\beta}(0)=\frac{8}{m}\left\langle\delta_{\alpha\beta}\left[2\hat{K}_\alpha+2\hat{V}_\alpha\right]+\frac{1}{2}\hat{U}\right\rangle_0.
\eq

Now, a {\sl collective excitation} at frequency $\omega_\alpha$ manifests itself with the appearance 
of a $\delta(\omega-\omega_\alpha)$ in the dissipation spectrum, i.e. 
${\rm Im}\chi_{\alpha\beta}(0,\omega)\propto\delta(\omega-\omega_\alpha)\delta_{\alpha\beta}$.

From the definition of the moments, Eq. (\ref{eq:Mnd}) follows
\bq
M^1_{\alpha\beta}&\propto&\omega_\alpha\delta_{\alpha\beta},\\
M^3_{\alpha\beta}&\propto&\omega_\alpha^3\delta_{\alpha\beta},
\eq
from which follows 
\bq
\left(\mlq M^3/M^1 \mrq\right)_{\alpha\beta}=\omega^2_\alpha\delta_{\alpha\beta}.
\eq
So we need to take the ``ratio'' $\mlq M^3/M^1 \mrq$ and eventually diagonalize the matrix. From Eqs. (\ref{eq:opE})-(\ref{eq:opT}) and using Feynman theorem we find
\bq \label{eq:FTo}
\left.\frac{\partial\langle\hat{H}_0\rangle_0}{\partial\omega_0^2}\right|_m=
\left\langle\left.\frac{\partial\hat{H}_0}{\partial\omega_0^2}\right|_m\right\rangle_0=
\frac{\langle\hat{V}_{ext}\rangle_0}{\omega_0^2},
\eq
from which follows $f_E=2f_V$. This is nothing else than the virial theorem again 
$2f_V=f_K+f_V+f_U$ or $f_K-f_V+f_U=0$ (which should be compared with Eq. (\ref{eq:vt}) which holds 
for $\alpha=1,2$). Here we use the same notation used in Eqs. (\ref{eq:opE})-(\ref{eq:opT}) with the
new interpretation that $f_E=\langle\hat{H}_0\rangle_0, f_K=\langle\hat{K}\rangle_0, f_V=\langle\hat{V}_{ext}\rangle_0, f_U=\langle\hat{U}\rangle_0$. Using Feynman theorem again, we 
also find
\bq \label{eq:FTm}
\left.\frac{\partial\langle\hat{H}_0\rangle_0}{\partial m}\right|_{\omega_0}=
\left\langle\left.\frac{\partial\hat{H}_0}{\partial m}\right|_{\omega_0}\right\rangle_0=
\frac{\langle\hat{V}_{ext}\rangle_0-\langle\hat{K}\rangle_0}{m},
\eq
from which follows $\partial f_E/\partial m=(f_V-f_K)/m$.

Using Eqs. (\ref{eq:FTo}) and (\ref{eq:FTm}) together
\bq
1-\frac{f_K}{f_V}=2\left(m\frac{\partial f_E}{\partial m}\frac{1}{f_E}\right)
=2\delta,
\eq
which defines $\delta$.

Let us now write $M^3/M^1$ using $f_K/f_V$. from Eqs. (\ref{eq:M1}) and (\ref{eq:M3}) follows
\bq 
M^1_{\alpha\beta}(0)&=&\delta_{\alpha\beta}\frac{4}{m\omega_0^2}\langle\hat{V}_{ext}\rangle_0,\\ \nonumber
M^3_{\alpha\beta}(0)&=&\frac{8}{m}\left\{\delta_{\alpha\beta}\left(2\langle\hat{K}_\alpha\rangle_0+2\langle\hat{V}_\alpha\rangle_0\right)+\frac{1}{2}\langle\hat{U}\rangle_0\right\}\\ \nonumber
&=&\frac{8}{m}\left\{\delta_{\alpha\beta}\left(\langle\hat{K}\rangle_0+\langle\hat{V}_{ext}\rangle_0\right)+\frac{1}{2}\langle\hat{U}\rangle_0\right\}\\ 
&=&\frac{4}{m}\left\{\delta_{\alpha\beta}2(\langle\hat{K}\rangle_0+\langle\hat{V}_{ext}\rangle_0)+(\langle\hat{V}_{ext}\rangle_0-\langle\hat{K}\rangle_0)\right\},
\eq
where for the third moment we used isotropy in the first equality and the virial theorem in the 
second. Upon rescaling $M^3$ 
\bq \nonumber
\left(\mlq M^3/M^1 \mrq\right)_{\alpha\beta}&=&\omega_0^2\left\{\delta_{\alpha\beta}2\left(\frac{f_K}{f_V}+1\right)+\left(1-\frac{f_K}{f_V}\right)\right\}\\ \nonumber
&=&\omega_0^2\left\{\delta_{\alpha\beta}2(2-2\delta)+2\delta\right\}\\
&=&\omega_0^2\left(\begin{array}{cc}4-2\delta & 2\delta\\ 2\delta & 4-2\delta\end{array}\right).
\eq
We can now diagonalize this matrix. The eigenvalues are
\bq \nonumber
4-2\delta-\lambda_\pm &=&\pm2\delta,\\
\lambda_\pm &=&4-2\delta\mp2\delta, 
\eq
and the eigenvectors ${\bf v}_\pm=(1,\pm 1)$. So that the monopole mode collective excitation has 
exactly a frequency
\bq
\omega_+=\sqrt{\lambda_+\omega_0^2}=\boxed{2\omega_0}.
\eq

\section{Conclusions}

In conclusion, we reviewed the linear response theory for quantum liquids with 
\red{fluctuation-dissipation theorem} and the associated 
Kramers-Kronig relations due to causality, we defined the longitudinal dielectric function for 
Coulomb liquids \red{setting a parallelism between the statistical mechanic properties of the 
many-body system and its electrostatics}. And we determined the monopole frequency for a \red{very 
degenerate, dilute} 
atomic \red{Bose} gas in a plane in a harmonic trap. \red{In particular we show that it coincides 
with twice the frequency of the trap}. The result we obtained through linear response theory is 
exact because we let the perturbation vanish in the end. \red{The result obtained is also 
independent from the parameters defining the Hamiltonian of the gas like its particles mass or 
the magnitude of their Dirac delta pair interaction. This can be important when studying trapped,
highly diluted, atomic (Bose) gases at very low temperatures.}

\appendix
\section{A particular commutator}
\label{app}

Given any two functions $f(\rr)$ and $g(\rr)$, we have
\bq \nonumber
&&[\int d^2\rr\,\hat{J}_\alpha(\rr)g(\rr),\int d^2\rr'\,\hat{\rho}(\rr')f(\rr')]=\\ \nonumber
&&-\frac{i\hbar}{m}\int d^2\rr\,g(\rr)\hat{\rho}(\rr)\nabla f(\rr)=\\ \label{eq:app}
&&\frac{i\hbar}{m}\int d^2\rr\,f(\rr)\nabla\left[g(\rr)\hat{\rho}(\rr)\right],
\eq
where the operators $\hat{\rho}$ and $\hat{J}_\alpha$ are defined in Eqs. (\ref{eq:rho}) and (\ref{eq:J}) of the main text and in the last equality we have used the fact that our fluid is bounded (i.e. the density 
operator decays to zero at $r\to\infty$) so that in the integration by parts we can neglect the 
surface term.

\section*{Author declarations}
\subsection*{Conflict of interest}
The author has no conflicts to disclose.

\section*{Data availability}
The data that support the findings of this study are available from the 
corresponding author upon reasonable request.
\bibliography{srql}

\begin{thebibliography}{21}%
\makeatletter
\providecommand \@ifxundefined [1]{%
 \@ifx{#1\undefined}
}%
\providecommand \@ifnum [1]{%
 \ifnum #1\expandafter \@firstoftwo
 \else \expandafter \@secondoftwo
 \fi
}%
\providecommand \@ifx [1]{%
 \ifx #1\expandafter \@firstoftwo
 \else \expandafter \@secondoftwo
 \fi
}%
\providecommand \natexlab [1]{#1}%
\providecommand \enquote  [1]{``#1''}%
\providecommand \bibnamefont  [1]{#1}%
\providecommand \bibfnamefont [1]{#1}%
\providecommand \citenamefont [1]{#1}%
\providecommand \href@noop [0]{\@secondoftwo}%
\providecommand \href [0]{\begingroup \@sanitize@url \@href}%
\providecommand \@href[1]{\@@startlink{#1}\@@href}%
\providecommand \@@href[1]{\endgroup#1\@@endlink}%
\providecommand \@sanitize@url [0]{\catcode `\\12\catcode `\$12\catcode
  `\&12\catcode `\#12\catcode `\^12\catcode `\_12\catcode `\%12\relax}%
\providecommand \@@startlink[1]{}%
\providecommand \@@endlink[0]{}%
\providecommand \url  [0]{\begingroup\@sanitize@url \@url }%
\providecommand \@url [1]{\endgroup\@href {#1}{\urlprefix }}%
\providecommand \urlprefix  [0]{URL }%
\providecommand \Eprint [0]{\href }%
\providecommand \doibase [0]{https://doi.org/}%
\providecommand \selectlanguage [0]{\@gobble}%
\providecommand \bibinfo  [0]{\@secondoftwo}%
\providecommand \bibfield  [0]{\@secondoftwo}%
\providecommand \translation [1]{[#1]}%
\providecommand \BibitemOpen [0]{}%
\providecommand \bibitemStop [0]{}%
\providecommand \bibitemNoStop [0]{.\EOS\space}%
\providecommand \EOS [0]{\spacefactor3000\relax}%
\providecommand \BibitemShut  [1]{\csname bibitem#1\endcsname}%
\let\auto@bib@innerbib\@empty
\bibitem [{\citenamefont {Pines}\ and\ \citenamefont
  {Nozi\`{e}res}(1966)}]{PinNoz66}%
  \BibitemOpen
  \bibfield  {author} {\bibinfo {author} {\bibfnamefont {D.}~\bibnamefont
  {Pines}}\ and\ \bibinfo {author} {\bibfnamefont {P.}~\bibnamefont
  {Nozi\`{e}res}},\ }\href@noop {} {\emph {\bibinfo {title} {The Teory of
  Quantum Liquids}}}\ (\bibinfo  {publisher} {W. A. Benjamin, Inc.},\ \bibinfo
  {address} {New York},\ \bibinfo {year} {1966})\ \bibinfo {note} {chapter
  2}\BibitemShut {NoStop}%
\bibitem [{\citenamefont {March}\ and\ \citenamefont {Tosi}(1984)}]{MarTos84}%
  \BibitemOpen
  \bibfield  {author} {\bibinfo {author} {\bibfnamefont {N.~H.}\ \bibnamefont
  {March}}\ and\ \bibinfo {author} {\bibfnamefont {M.~P.}\ \bibnamefont
  {Tosi}},\ }\href@noop {} {\emph {\bibinfo {title} {Coulomb Liquids}}}\
  (\bibinfo  {publisher} {Academic Press},\ \bibinfo {address} {London},\
  \bibinfo {year} {1984})\ \bibinfo {note} {appendix 5}\BibitemShut {NoStop}%
\bibitem [{\citenamefont {Hansen}\ and\ \citenamefont
  {McDonald}(1986)}]{HanMcD86}%
  \BibitemOpen
  \bibfield  {author} {\bibinfo {author} {\bibfnamefont {J.~P.}\ \bibnamefont
  {Hansen}}\ and\ \bibinfo {author} {\bibfnamefont {I.~R.}\ \bibnamefont
  {McDonald}},\ }\href@noop {} {\emph {\bibinfo {title} {Theory of Simple
  Liquids}}},\ \bibinfo {edition} {2nd}\ ed.\ (\bibinfo  {publisher} {Academic
  Press},\ \bibinfo {address} {London},\ \bibinfo {year} {1986})\ \bibinfo
  {note} {section 7.6}\BibitemShut {NoStop}%
\bibitem [{\citenamefont {Chaikin}\ and\ \citenamefont
  {Lubensky}(2000)}]{Chaikin-Lubensky}%
  \BibitemOpen
  \bibfield  {author} {\bibinfo {author} {\bibfnamefont {P.~M.}\ \bibnamefont
  {Chaikin}}\ and\ \bibinfo {author} {\bibfnamefont {T.~C.}\ \bibnamefont
  {Lubensky}},\ }\href@noop {} {\emph {\bibinfo {title} {{Principles of
  Condensed Matter Physics}}}}\ (\bibinfo  {publisher} {Cambridge University
  Press},\ \bibinfo {year} {2000})\BibitemShut {NoStop}%
\bibitem [{\citenamefont {Giuliani}\ and\ \citenamefont
  {Vignale}(2005)}]{Giuliani-Vignale}%
  \BibitemOpen
  \bibfield  {author} {\bibinfo {author} {\bibfnamefont {G.}~\bibnamefont
  {Giuliani}}\ and\ \bibinfo {author} {\bibfnamefont {G.}~\bibnamefont
  {Vignale}},\ }\href@noop {} {\emph {\bibinfo {title} {{Quantum Theory of the
  Electron Liquid}}}}\ (\bibinfo  {publisher} {Cambridge University Press},\
  \bibinfo {year} {2005})\BibitemShut {NoStop}%
\bibitem [{\citenamefont {Pathria}\ and\ \citenamefont
  {Beale}(2021)}]{Pathria-Beale}%
  \BibitemOpen
  \bibfield  {author} {\bibinfo {author} {\bibfnamefont {R.~K.}\ \bibnamefont
  {Pathria}}\ and\ \bibinfo {author} {\bibfnamefont {P.~D.}\ \bibnamefont
  {Beale}},\ }\href@noop {} {\emph {\bibinfo {title} {{Statistical
  Mechanics}}}}\ (\bibinfo  {publisher} {Academic Press},\ \bibinfo {year}
  {2021})\BibitemShut {NoStop}%
\bibitem [{\citenamefont {Fantoni}\ and\ \citenamefont
  {Tosi}(1995{\natexlab{a}})}]{Fantoni95b}%
  \BibitemOpen
  \bibfield  {author} {\bibinfo {author} {\bibfnamefont {R.}~\bibnamefont
  {Fantoni}}\ and\ \bibinfo {author} {\bibfnamefont {M.~P.}\ \bibnamefont
  {Tosi}},\ }\bibfield  {title} {\bibinfo {title} {{Coordinate space form of
  interacting reference response function of d-dimensional jellium}},\ }\href
  {https://doi.org/10.1007/BF02454131} {\bibfield  {journal} {\bibinfo
  {journal} {Nuovo Cimento}\ }\textbf {\bibinfo {volume} {17D}},\ \bibinfo
  {pages} {1165} (\bibinfo {year} {1995}{\natexlab{a}})}\BibitemShut {NoStop}%
\bibitem [{\citenamefont {Fantoni}(2013)}]{Fantoni13g}%
  \BibitemOpen
  \bibfield  {author} {\bibinfo {author} {\bibfnamefont {R.}~\bibnamefont
  {Fantoni}},\ }\bibfield  {title} {\bibinfo {title} {{Radial distribution
  function in a diffusion Monte Carlo simulation of a Fermion fluid between the
  ideal gas and the Jellium model}},\ }\href
  {https://doi.org/10.1140/epjb/e2013-40204-3} {\bibfield  {journal} {\bibinfo
  {journal} {Eur. Phys. J. B}\ }\textbf {\bibinfo {volume} {86}},\ \bibinfo
  {pages} {286} (\bibinfo {year} {2013})}\BibitemShut {NoStop}%
\bibitem [{\citenamefont {Fantoni}(2018)}]{Fantoni18c}%
  \BibitemOpen
  \bibfield  {author} {\bibinfo {author} {\bibfnamefont {R.}~\bibnamefont
  {Fantoni}},\ }\bibfield  {title} {\bibinfo {title} {{One-component fermion
  plasma on a sphere at finite temperature}},\ }\href
  {https://doi.org/10.1142/S012918311850064X} {\bibfield  {journal} {\bibinfo
  {journal} {Int. J. Mod. Phys. C}\ }\textbf {\bibinfo {volume} {29}},\
  \bibinfo {pages} {1850064} (\bibinfo {year} {2018})}\BibitemShut {NoStop}%
\bibitem [{\citenamefont {Fantoni}(2021{\natexlab{a}})}]{Fantoni21b}%
  \BibitemOpen
  \bibfield  {author} {\bibinfo {author} {\bibfnamefont {R.}~\bibnamefont
  {Fantoni}},\ }\bibfield  {title} {\bibinfo {title} {{Jellium at finite
  temperature using the restricted worm algorithm}},\ }\href
  {https://doi.org/10.1140/epjb/s10051-021-00078-y} {\bibfield  {journal}
  {\bibinfo  {journal} {Eur. Phys. J. B}\ }\textbf {\bibinfo {volume} {94}},\
  \bibinfo {pages} {63} (\bibinfo {year} {2021}{\natexlab{a}})}\BibitemShut
  {NoStop}%
\bibitem [{\citenamefont {Fantoni}(2021{\natexlab{b}})}]{Fantoni21d}%
  \BibitemOpen
  \bibfield  {author} {\bibinfo {author} {\bibfnamefont {R.}~\bibnamefont
  {Fantoni}},\ }\bibfield  {title} {\bibinfo {title} {{Form invariance of the
  moment sum-rules for jellium with the addition of short-range terms in the
  pair-potential}},\ }\href {https://doi.org/10.1007/s12648-020-01750-2}
  {\bibfield  {journal} {\bibinfo  {journal} {Indian J. Phys.}\ }\textbf
  {\bibinfo {volume} {95}},\ \bibinfo {pages} {1027} (\bibinfo {year}
  {2021}{\natexlab{b}})}\BibitemShut {NoStop}%
\bibitem [{\citenamefont {Fantoni}(2021{\natexlab{c}})}]{Fantoni21i}%
  \BibitemOpen
  \bibfield  {author} {\bibinfo {author} {\bibfnamefont {R.}~\bibnamefont
  {Fantoni}},\ }\bibfield  {title} {\bibinfo {title} {{Jellium at finite
  temperature}},\ }\href {https://doi.org/10.1080/00268976.2021.1996648}
  {\bibfield  {journal} {\bibinfo  {journal} {Mol. Phys.}\ }\textbf {\bibinfo
  {volume} {120}},\ \bibinfo {pages} {4} (\bibinfo {year}
  {2021}{\natexlab{c}})}\BibitemShut {NoStop}%
\bibitem [{\citenamefont {Fantoni}\ and\ \citenamefont
  {Tosi}(1995{\natexlab{b}})}]{Fantoni95a}%
  \BibitemOpen
  \bibfield  {author} {\bibinfo {author} {\bibfnamefont {R.}~\bibnamefont
  {Fantoni}}\ and\ \bibinfo {author} {\bibfnamefont {M.~P.}\ \bibnamefont
  {Tosi}},\ }\bibfield  {title} {\bibinfo {title} {{Decay of correlations and
  related sum rules in a layered classical plasma}},\ }\href
  {https://doi.org/10.1007/BF02451594} {\bibfield  {journal} {\bibinfo
  {journal} {Nuovo Cimento}\ }\textbf {\bibinfo {volume} {17D}},\ \bibinfo
  {pages} {155} (\bibinfo {year} {1995}{\natexlab{b}})}\BibitemShut {NoStop}%
\bibitem [{\citenamefont {Fantoni}\ \emph {et~al.}(2003)\citenamefont
  {Fantoni}, \citenamefont {Jancovici},\ and\ \citenamefont
  {T\'ellez}}]{Fantoni03a}%
  \BibitemOpen
  \bibfield  {author} {\bibinfo {author} {\bibfnamefont {R.}~\bibnamefont
  {Fantoni}}, \bibinfo {author} {\bibfnamefont {B.}~\bibnamefont {Jancovici}},\
  and\ \bibinfo {author} {\bibfnamefont {G.}~\bibnamefont {T\'ellez}},\
  }\bibfield  {title} {\bibinfo {title} {{Pressures for a One-Component Plasma
  on a Pseudosphere}},\ }\href {https://doi.org/10.1023/A:1023671419021}
  {\bibfield  {journal} {\bibinfo  {journal} {J. Stat. Phys.}\ }\textbf
  {\bibinfo {volume} {112}},\ \bibinfo {pages} {27} (\bibinfo {year}
  {2003})}\BibitemShut {NoStop}%
\bibitem [{\citenamefont {Fantoni}\ and\ \citenamefont
  {T\'ellez}(2008)}]{Fantoni08c}%
  \BibitemOpen
  \bibfield  {author} {\bibinfo {author} {\bibfnamefont {R.}~\bibnamefont
  {Fantoni}}\ and\ \bibinfo {author} {\bibfnamefont {G.}~\bibnamefont
  {T\'ellez}},\ }\bibfield  {title} {\bibinfo {title} {{Two dimensional
  one-component plasma on a Flamm's paraboloid}},\ }\href
  {https://doi.org/10.1007/s10955-008-9616-x} {\bibfield  {journal} {\bibinfo
  {journal} {J. Stat. Phys.}\ }\textbf {\bibinfo {volume} {133}},\ \bibinfo
  {pages} {449} (\bibinfo {year} {2008})}\BibitemShut {NoStop}%
\bibitem [{\citenamefont {Fantoni}(2012)}]{Fantoni12b}%
  \BibitemOpen
  \bibfield  {author} {\bibinfo {author} {\bibfnamefont {R.}~\bibnamefont
  {Fantoni}},\ }\bibfield  {title} {\bibinfo {title} {{Two component plasma in
  a Flamm's paraboloid}},\ }\href
  {https://doi.org/10.1088/1742-5468/2012/04/P04015} {\bibfield  {journal}
  {\bibinfo  {journal} {J. Stat. Mech.}\ ,\ \bibinfo {pages} {P04015}}
  (\bibinfo {year} {2012})}\BibitemShut {NoStop}%
\bibitem [{\citenamefont {Fantoni}\ and\ \citenamefont
  {Pastore}(2013{\natexlab{a}})}]{Fantoni13e}%
  \BibitemOpen
  \bibfield  {author} {\bibinfo {author} {\bibfnamefont {R.}~\bibnamefont
  {Fantoni}}\ and\ \bibinfo {author} {\bibfnamefont {G.}~\bibnamefont
  {Pastore}},\ }\bibfield  {title} {\bibinfo {title} {{The restricted primitive
  model of ionic fluids with nonadditive diameters}},\ }\href
  {https://doi.org/10.1209/0295-5075/101/46003} {\bibfield  {journal} {\bibinfo
   {journal} {Europhys. Lett.}\ }\textbf {\bibinfo {volume} {101}},\ \bibinfo
  {pages} {46003} (\bibinfo {year} {2013}{\natexlab{a}})}\BibitemShut {NoStop}%
\bibitem [{\citenamefont {Fantoni}\ and\ \citenamefont
  {Pastore}(2013{\natexlab{b}})}]{Fantoni13f}%
  \BibitemOpen
  \bibfield  {author} {\bibinfo {author} {\bibfnamefont {R.}~\bibnamefont
  {Fantoni}}\ and\ \bibinfo {author} {\bibfnamefont {G.}~\bibnamefont
  {Pastore}},\ }\bibfield  {title} {\bibinfo {title} {{Monte Carlo simulation
  of the nonadditive restricted primitive model of ionic fluids: Phase diagram
  and clustering}},\ }\href {https://doi.org/10.1103/PhysRevE.87.052303}
  {\bibfield  {journal} {\bibinfo  {journal} {Phys. Rev. E}\ }\textbf {\bibinfo
  {volume} {87}},\ \bibinfo {pages} {052303} (\bibinfo {year}
  {2013}{\natexlab{b}})}\BibitemShut {NoStop}%
\bibitem [{\citenamefont {Alastuey}\ and\ \citenamefont
  {Fantoni}(2016)}]{Fantoni16b}%
  \BibitemOpen
  \bibfield  {author} {\bibinfo {author} {\bibfnamefont {A.}~\bibnamefont
  {Alastuey}}\ and\ \bibinfo {author} {\bibfnamefont {R.}~\bibnamefont
  {Fantoni}},\ }\bibfield  {title} {\bibinfo {title} {{Fourth moment sum rule
  for the charge correlations of a two-component classical plasma}},\ }\href
  {https://doi.org/10.1007/s10955-016-1512-1} {\bibfield  {journal} {\bibinfo
  {journal} {J. Stat. Phys.}\ }\textbf {\bibinfo {volume} {163}},\ \bibinfo
  {pages} {887} (\bibinfo {year} {2016})}\BibitemShut {NoStop}%
\bibitem [{\citenamefont {Fantoni}(2019)}]{Fantoni19a}%
  \BibitemOpen
  \bibfield  {author} {\bibinfo {author} {\bibfnamefont {R.}~\bibnamefont
  {Fantoni}},\ }\bibfield  {title} {\bibinfo {title} {{Plasma living in a
  curved surface at some special temperature}},\ }\href
  {https://doi.org/10.1016/j.physa.2019.04.222} {\bibfield  {journal} {\bibinfo
   {journal} {Physica A}\ }\textbf {\bibinfo {volume} {177}},\ \bibinfo {pages}
  {524} (\bibinfo {year} {2019})}\BibitemShut {NoStop}%
\bibitem [{\citenamefont {Fantoni}(2023)}]{Fantoni23a}%
  \BibitemOpen
  \bibfield  {author} {\bibinfo {author} {\bibfnamefont {R.}~\bibnamefont
  {Fantoni}},\ }\bibfield  {title} {\bibinfo {title} {{One-component fermion
  plasma on a sphere at finite temperature. The anisotropy in the paths
  conformations}},\ }\href {https://doi.org/10.1088/1742-5468/aceb54}
  {\bibfield  {journal} {\bibinfo  {journal} {J. Stat. Mech.}\ ,\ \bibinfo
  {pages} {083103}} (\bibinfo {year} {2023})}\BibitemShut {NoStop}%
\end{thebibliography}%

\end{document}